\documentclass[reprint, superscriptaddress,preprintnumbers,nofootinbib,aps]{revtex4-1}
\pdfoutput=1

\usepackage{graphicx}
\usepackage{amsmath,amssymb,amsfonts,amsthm,stmaryrd,mathtools,mathbbol,bm,physics,tensor}
\usepackage{soul}
\usepackage{xcolor}
\usepackage{tikz}
\allowdisplaybreaks[1]
\usepackage[bookmarks,linktocpage, colorlinks=true, plainpages = false, citecolor = treegreen,  linkcolor=darkblue, urlcolor = darkblue, filecolor = blue]{hyperref} 

\definecolor{tealgreen}{rgb}{0.0, 0.5, 1.0}
\definecolor{darkblue}{rgb}{0., 0.4, 0.8}
\definecolor{cadmiumred}{rgb}{1., 0., 0.22}
\definecolor{treegreen}{rgb}{0., 0.7, 0.3}

\usepackage{pifont}

\def\be#1\ee{\begin{align}#1\end{align}}

\def\ba{\begin{eqnarray}}
\def\ea{\end{eqnarray}}
\def\nn{\nonumber}

\usepackage[normalem]{ulem}

\begin{document}

\title{$g_{tt}g_{rr} =-1$ black hole thermodynamics
in general quasi-topological gravity
}

\author{Johanna Borissova}
\email{j.borissova@imperial.ac.uk}
\affiliation{Abdus Salam Centre for Theoretical Physics, Imperial College London, London SW7 2AZ, UK}

\begin{abstract}
\bigskip
{\sc Abstract:} We 
present a 
unified framework for the discussion of black hole thermodynamics of $d$-dimensional static black holes with spherical or compact hyperbolic horizon topology satisfying $g_{tt}g_{rr}=-1$ in Schwarzschild gauge. To that end, we consider any such black hole as a solution to an integrable  $2$-dimensional effective dilaton theory and thereby as a vacuum solution to a generalised notion of $d$-dimensional quasi-topological gravity. We show that the generating function determining $f(r) = -g_{tt} $ in the integrated equation of motion
provides the thermodynamic mass in a generalised first law with entropy computed as the Wald entropy. The framework presented here can be applied to singular and regular black holes with flat or anti-de Sitter asymptotics.
~\\

\noindent

\end{abstract}

\maketitle


\bigskip

The Bekenstein-Hawking area law for the entropy of a black hole in general relativity (GR)~\cite{Bekenstein:1973ur,Hawking:1975vcx} is expected to receive corrections
for beyond-GR black holes. These corrections can be accessed  through Wald's Noether charge formalism~\cite{Wald:1993nt,Iyer:1994ys} --- provided a given beyond-GR black hole arises as a known solution to a generally covariant gravitational theory. This is often not the case for phenomenological black hole models inspired by
 quantum gravity.~\footnote{
 See e.g.~\cite{Bonanno:2000ep,Bonanno:2006eu,Falls:2010he,Koch:2013owa,Pawlowski:2018swz,Adeifeoba:2018ydh,Platania:2019kyx,Eichhorn:2022bgu,Platania:2023srt,Fragomeno:2024tlh,Borissova:2022mgd,Bonanno:2024wvb} for a non-exhaustive set of quantum gravity inspired examples related to regular black hole models such as those proposed originally in~\cite{Bardeen:1968bh,Dymnikova:1992ux,Hayward:2005gi}. Here we are not discussing the case when such models are produced by coupling GR to exotic matter violating energy conditions, such as for example non-linear electrodynamic sources~\cite{Ayon-Beato:1998hmi,Ayon-Beato:1999qin,Ayon-Beato:1999kuh,Ayon-Beato:2000mjt,Bronnikov:2000vy,Dymnikova:2004zc,Bronnikov:2022ofk,Balart:2014cga,Fan:2016hvf,Rodrigues:2018bdc,AraujoFilho:2026sna}.} 
In this case dealing with the thermodynamics of black holes typically requires imposing the first law by hand. The purpose of this letter is to establish a unified framework for the discussion of black hole thermodynamics of $d\geq 4 $ dimensional static black holes satisfying $g_{tt}g_{rr}=-1$ in Schwarzschild gauge as gravitational vacuum solutions to a generalised notion of $d$-dimensional {\it quasi-topological gravity} (QTG)
which allows us to ground the discussion on integrable $2$-dimensional generalised dilaton  theories.~\footnote{See~\cite{Banks:1990mk,Louis-Martinez:1993bge} for the early stages of development of $2$-dimensional dilaton theory and e.g.~\cite{Grumiller:2002nm} for a review.}

The precise albeit {\it implicit} definition~\cite{Borissova:2026krh} of a {\it general} QTG adopted here is detailed further below and can be summarised as follows. A general QTG is {\it any} $d$-dimensional gravitational theory whose reduction on spherically symmetric backgrounds --- or backgrounds with compact hyperbolic or toroidal symmetry --- yields a $2$-dimensional Horndeski theory satisfying an integrability condition which is necessary and sufficient for vacuum solutions to be static and to satisfy $g_{tt}g_{rr} = -1$ in Schwarzschild gauge, whereby the first-order equation determining $g_{tt}=-f(r)$ can be integrated into an algebraic equation involving a theory-dependent generating function $\Omega(r,f)$.~\footnote{The exception to this statement are theories for which $\partial_f \Omega(r,f) = 0$ onshell~\cite{Borissova:2026krh}.} We will henceforth refer to the reduced theory as a $2$-dimensional {\it integrable} Horndeski theory. Conversely, lifting such a $2$-dimensional integrable Horndeski theory to a $d\geq 4$ dimensional gravity~\cite{Borissova:2026krh,Borissova:2026wmn,Colleaux:2019ckh} by this definition produces a general QTG. This definition is operational in what concerns the derivation of spherically symmetric vacuum solutions, and does not require or  assume knowledge about the precise form of the $d$-dimensional gravitational action.

 	We will carefully distinguish this generalised notion of a QTG from the following notions of QTGs so far appearing in the literature. The first one are QTGs as an {\it explicit} set of theories representing polynomial higher-curvature extensions of general relativity~\cite{Oliva:2010eb,Myers:2010jv,Dehghani:2011vu,Cisterna:2017umf,Bueno:2019ltp,Bueno:2019ycr, Bueno:2022res,Moreno:2023rfl,Bueno:2025qjk} whose reduction on spherically symmetric backgrounds produces a subset of $2$-dimensional integrable  Horndeski theories~\cite{Bueno:2024zsx,Bueno:2024eig,Bueno:2025qjk,Bueno:2025gjg,Borissova:2026krh}.
 	We will refer to such theories here explicitly as {\it polynomial curvature} QTGs. 
 	These are contained  in the larger set of QTGs built from arbitrary not necessarily analytic functions of curvature invariants without covariant derivatives~\cite{Bueno:2025zaj,Borissova:2026wmn,Borissova:2026krh}. We will refer here to these latter theories as {\it non-polynomial curvature} QTGs.
 	General QTGs contain polynomial and non-polynomial curvature QTGs as a subclass, but represent an extended class of theories satisfying the  above defining properties with an action containing in general arbitrary functions of curvature and curvature-derivative invariants~\cite{Borissova:2026krh}. 
 	This is the relevant class of theories on which our discussion is grounded.

The notion of a general QTG is not to be confused with the notion of a {\it generalised} QTG (GQTG) used in previous works~\cite{Bueno:2019ltp,Bueno:2019ycr, Bueno:2022res,Moreno:2023rfl} to refer to gravitational theories representing polynomial higher-curvature extensions of general relativity which admit static spherically symmetric vacuum solutions satisfying $g_{tt}g_{rr} = -1$, whereby the equation determining $g_{tt}=-f(r)$  is algebraic {\it only} if the GQTG is a polynomial curvature QTG. 
The reduction on spherically symmetric backgrounds of a GQTG which is not a polynomial curvature QTG thus cannot be a $2$-dimensional integrable Horndeski theory.\\

{\it General quasi-topological gravities---}Consider a $d$-dimensional static spacetime
\ba\label{eq:Metricf}
\dd{s}^2 &=& -f(r)\dd{t}^2 + \frac{\dd{r}^2}{f(r)} + r^2 \dd{\Sigma^2_{d-2}}\quad
\ea
where $\dd{\Sigma^2_{d-2}} $ is the surface element of a $d-2$ dimensional compact space of constant curvature $k = \pm 1$ corresponding to spherical or compact hyberbolic symmetry. Any such spacetime depending on a constant dimensionful parameter
can be constructed as a vacuum solution to a
general QTG~\cite{Borissova:2026krh}.~\footnote{Extending this statement in full generality to spacetimes~\eqref{eq:Metricf} with $k=0$ would require finding covariant densities which separate the variables $\varphi$ and $\chi$ for a general warped-product metric~\eqref{eq:MetricWarped} as done for $k \neq 0 $ in~\cite{Borissova:2026krh}. 
It is not clear whether such densities exist.
The thermodynamic relations derived here nevertheless apply to spacetimes~\eqref{eq:Metricf} with $k=0$ if the generating function $\Omega(r,f)$ determined by the inversion of $f(r) $ with respect to the dimensionful parameter --- as discussed later --- takes the particular form $\Omega(r,f) \propto r^{d-1}h\qty(\frac{f}{r^2})$, where $h$ is an arbitrary function of its argument. These spacetimes can be constructed as vacuum solutions to a general QTG~\cite{Borissova:2026krh,Borissova:2026wmn}.} 

By a general QTG we mean here
a generally covariant theory~\footnote{We work in units where $c = \hbar = k_{\text{B}} =1$ and Newton's constant has mass dimension $\qty[G_\text{N}] = 2-d$.} 
\ba \label{eq:S}
S[g] &=& \frac{1}{16 \pi G_\text{N}}\int \dd[d]{x} \sqrt{-g} %
\,\mathcal{L}_g,\quad\\
\mathcal{L}_g & = & -2\Lambda + R + \Delta \mathcal{L}_g\label{eq:Lg},
\ea
with second-order equations on warped-product backgrounds
\ba\label{eq:MetricWarped}
g_{\mu\nu}(x)\dd{x}^\mu \dd{x}^\nu & =& q_{ab}(y)\dd{y}^a \dd{y}^b + \varphi(y)^2 \dd{\Sigma_{d-2}^2},\quad \\
\dd{\Sigma^2_{d-2}} &=& \gamma_{ij}(\theta)\dd{\theta}^i \dd{\theta}^j,
\ea
for which the $2$-dimensional reduced
 theory
\ba\label{eq:SHorndeski}
S_{\text{2D}}[q,\varphi]  & =&  \frac{(d-2)\Sigma_{d-2}}{16 \pi G_\text{N}}
  \int \dd[2]{y} \sqrt{-q}\,   \mathcal{L}_{q,\varphi},\quad \\
 \mathcal{L}_{q,\varphi} & = & \frac{1}{d-2} \varphi^{d-2} \eval{\mathcal{L}_g}_{\eqref{eq:MetricWarped}},\quad
 \ea
 satisfies an integrability condition which is necessary and sufficient for the existence of solutions~\eqref{eq:Metricf} and the equation of motion for $f$ to be integrable into an algebraic equation. The generating function determining $f$ for a black hole can then be interpreted as the thermodynamic mass,
 as we will discuss here.

Since the reduction of a general QTG on~\eqref{eq:MetricWarped}, by definition, is a $2$-dimensional  Horndeski theory for $q_{ab}$ and $\varphi$, the Lagrangian density of the reduced action~\eqref{eq:SHorndeski} takes the form~\cite{Horndeski:1974wa,Kobayashi:2011nu}
 \ba\label{eq:L2D}
 \mathcal{L}_{q,\varphi} &=&
h_2(\varphi,\chi) - h_3(\varphi,\chi)\Box \varphi
+ h_4(\varphi,\chi) {\mathcal{R}}\nn\\
&+& 2 \partial_\chi h_4(\varphi,\chi)\big[\qty(\Box \varphi)^2 -\nabla_a  \nabla_b \varphi \nabla^a \nabla^b \varphi\big],\,\quad 
\ea
where $\nabla_a$ is the covariant derivative associated to $q_{ab}$, $\mathcal{R} $ its Ricci scalar, and $h_i$ for $i =2,3,4$ are theory-dependent functions of the scalar field $\varphi$ and its kinetic term $\chi = \nabla_a \varphi \nabla^a \varphi$. The  former is considered as a field with mass dimension $[\varphi] = -1$ in the reduced action~\eqref{eq:SHorndeski}.

The integrability condition satisfied by the reduced action~\eqref{eq:SHorndeski} of a general QTG is~\cite{Borissova:2026krh,Borissova:2026wmn}
\ba\label{eq:Integrability}
\partial_\chi \alpha(\varphi,\chi) - \partial_\varphi \beta(\varphi,\chi)  &=& 0, \quad 
\ea
where
\ba
\alpha(\varphi,\chi)  &=&   h_2 + \chi \partial_\varphi \qty( h_3  - 2 \partial_\varphi h_4),\label{eq:Alpha}\quad \\
\beta(\varphi,\chi)  &=&  \chi \partial_\chi \qty(h_3  - 2 \partial_\varphi h_4) - \partial_\varphi h_4,\label{eq:Beta} \quad 
\ea
are two functions with mass dimensions $\qty[\alpha] = 4-d$ and $\qty[\beta] = 3-d$ which determine the equations of motion for $q_{ab}$ and $\varphi$.\\

The implications of the integrability condition~\eqref{eq:Integrability} can be seen as follows. According to the principle of symmetric criticality which holds for compact symmetry groups~\cite{Palais:1979rca,Fels:2001rv,Deser:2003up,Torre:2010xa,Frausto:2024egp}, the field equations derived from the generally covariant action~\eqref{eq:S} and evaluated on the ansatz~\eqref{eq:MetricWarped} are equivalent to the field equations derived from the reduced action~\eqref{eq:SHorndeski}. The corresponding variations are related by
\ba
\eval{\mathcal{E}_{\mu\nu}}_{\eqref{eq:MetricWarped}} 
& \vcentcolon = & \eval{\frac{16 \pi G_\text{N}}{\sqrt{-g} } \frac{\var S}{\var g^{\mu\nu}}}_{\eqref{eq:MetricWarped}} \nn\\
&=&
(d-2)\varphi^{2-d} \mathcal{E}_{ab} \delta^{a}_\mu \delta^b_\nu 
- \frac{1}{2
} \varphi^{5-d} \mathcal{E}_\varphi \gamma_{ij} \delta^i_\mu \delta^j_\nu ,\,\,\,\quad\,\,\label{eq:Emunu}
\ea
where 
\ba
\mathcal{E}_{ab} & \vcentcolon = & \frac{16 \pi G_\text{N}}{(d-2)\Sigma_{d-2}} \frac{1}{\sqrt{-q} }\frac{\var S_{\text{2D}}}{\var q^{ab}},\quad \,\,\,\,\\
\mathcal{E}_\varphi & \vcentcolon = & \frac{16 \pi G_\text{N}}{(d-2)\Sigma_{d-2} } \frac{1}{\sqrt{-q} } \frac{\var S_{\text{2D}}}{\var \varphi }.\quad \,\,\,\,
\ea
When the integrability condition~\eqref{eq:Integrability} is satisfied, the equation of motion $\mathcal{E}_{ab}=0$ for $q_{ab}$ is explicitly~\cite{Boyanov:2025pes,Carballo-Rubio:2025ntd}
\ba\label{eq:Eab}
\frac{1}{2}\big(\alpha + 2\beta \Box \varphi\big)q_{ab}  -
  \beta \nabla_a \nabla_b \varphi &=& 0, \quad \,
\ea
whereas the equation of motion $\mathcal{E}_\varphi =0$ for $\varphi$ is redundant onshell on the equation of motion for $q_{ab}$ due to the offshell Bianchi identity
\ba\label{eq:Bianchi}
\nabla^a \mathcal{E}_{ab} + \frac{1}{2} \mathcal{E}_\varphi \nabla_b \varphi &=& 0
\ea
resulting from $2$-dimensional general covariance of~\eqref{eq:SHorndeski} provided the scalar field is non-constant.
This is equivalent to the statement that the angular components of the equation of motion $\mathcal{E}_{\mu\nu} = 0$ for $g_{\mu\nu}$ in~\eqref{eq:MetricWarped} are redundant due to the offshell Bianchi identity $\nabla^\mu \mathcal{E}_{\mu\nu} = 0$ resulting from $d$-dimensional general covariance of~\eqref{eq:S}.
For an ansatz
\ba
q_{ab} \qty(y)\dd{y}^a \dd{y}^b  &=& -n(t,r)^2 f(t,r)\dd{t}^2 + \frac{\dd{r}^2}{f(t,r)},\quad\,\,\, \label{eq:q}\\
\varphi(y) &=& r\label{eq:Phi},
\ea
i.e., $\chi = f$,
 the independent components of the equation of motion~\eqref{eq:Eab} yield the three equations
\ba\label{eq:EQAll}
\partial_t f \,\,=\,\, 0, \quad \partial_r n \,\,= \,\, 0,\quad  \alpha + \beta \partial_r f &=& 0,\quad \,\,
\ea
where $\beta \neq 0$ is assumed.
A general QTG thus satisfies the following Birkhoff theorem~\cite{Borissova:2026krh} --- For generic solutions of the form
\ba\label{eq:gSpherical}
\dd{s}^2 &=& -n(t,r)^2 f(t,r)\dd{t}^2 + \frac{\dd{r}^2}{f(t,r)} + r^2 \dd{\Sigma^2_{d-2}}\,\,\quad \,\,
\ea
 in vacuum, the metric  function $f$ is time-independent and one may set $n=1$ after a redefinition of the coordinate $t$ such that the solutions are static and satisfy $g_{tt}g_{rr} = -1$. The integrability condition~\eqref{eq:Integrability} moreover implies that the remaining equation determining $f=f(r)$ in~\eqref{eq:EQAll} can be written in terms of a generating function $\Omega$ satisfying the defining offshell relations~\cite{Boyanov:2025pes,Carballo-Rubio:2025ntd}
\ba
\alpha(\varphi,\chi)&=& \frac{16 \pi G_\text{N} }{(d-2)\Sigma_{d-2}}\partial_\varphi \Omega(\varphi,\chi), \label{eq:AlphaBeta1}\quad \\
 \beta(\varphi,\chi) &=& \frac{16 \pi G_\text{N} }{(d-2)\Sigma_{d-2}}  \partial_\chi \Omega(\varphi,\chi),\label{eq:AlphaBeta2}\,\,\, \quad 
\ea
thus having mass dimension $\qty[\Omega] = 1$,
as 
\ba\label{eq:OmegaEOM}
\derivative{}{r} \Omega(r,f) &=& \partial_r \Omega(r,f) + \partial_f \Omega(r,f) f' \,\,= \,\, 0,\quad \quad 
\ea
and can be integrated into an algebraic equation for $f$,
\ba\label{eq:Omega2M}
\Omega(r,f) &=& 
M,\quad 
\ea
where $M$ is an integration constant. 
We will see that this identification provides the thermodynamic
 mass in a first law for $d$-dimensional static black holes~\eqref{eq:Metricf}.\\

General QTGs include Lovelock theories~\cite{Lanczos:1938sf,Lovelock:1970zsf,Lovelock:1971yv} and polynomial curvature QTGs~\cite{Oliva:2010eb,Myers:2010jv,Dehghani:2011vu,Cisterna:2017umf,Bueno:2019ltp,Bueno:2019ycr, Bueno:2022res,Moreno:2023rfl}  satisfying an analogue version of the above Birkhoff theorem~\cite{Bueno:2025qjk}. For these theories the generating function $\Omega$ takes the form~\cite{Bueno:2024dgm,Hennigar:2025yqm,Aguayo:2025xfi} 
\ba\label{eq:OmegaPolynomialCQTG}
\Omega(r,f) \,\,=\,\, \frac{(d-2)\Sigma_{d-2}}{16 \pi G_\text{N}} r^{d-1} h\qty(\frac{k-f}{r^2}), \quad 
\ea
where $h$ is a power series in its argument. The reduction of GR with a negative cosmological constant
is described by the function
\ba
h\qty(\frac{k-f}{r^2}) &=&- \frac{2 \Lambda}{(d-1)(d-2)} +\frac{k-f}{r^2} \label{eq:H1}\quad \quad \\
&=\vcentcolon&  \frac{1}{\ell^2} +\frac{k-f}{r^2},\label{eq:H2}
\ea
where $\ell$ is the anti-de Sitter radius.
The integrated equation of motion~\eqref{eq:Omega2M} in this case yields
\ba\label{eq:fGR}
f(r) &=& k + \frac{r^2}{\ell^2} - \frac{2m}{r^{d-3}},\quad 
\ea
where the parameter $m$ is related to the  integration constant $M$ by
\ba
M &=& \frac{(d-2)\Sigma_{d-2}}{16 \pi G_{\text{N}}}\,2m.
\ea
Thus $M$ is identical to the Arnowitt-Deser-Misner mass~\cite{Arnowitt:1959ah,Arnowitt:1962hi} of the spacetime.

Polynomial curvature QTGs beyond GR exist only in $d\geq 5$ dimensions~\cite{Bueno:2019ltp,Moreno:2023rfl}, whereas
 non-polynomial curvature QTGs can be constructed also in $d=4$ dimensions~\cite{Borissova:2026wmn,Borissova:2026krh,Bueno:2025zaj,Colleaux:2017ibe} and in this case $\Omega$ can have a more general dependence on $r$ and $f$ than in~\eqref{eq:OmegaPolynomialCQTG}~\cite{Borissova:2026wmn,Borissova:2026krh}.~\footnote{For applications of $d\geq 5$ polynomial and $d=4$ non-polynomial curvature QTGs in the context of black holes and cosmology, see e.g.~\cite{Oliva:2010eb,Myers:2010jv,Dehghani:2011vu,Cisterna:2017umf,Bueno:2019ltp,Bueno:2019ycr, Bueno:2022res,Moreno:2023rfl,Bueno:2024dgm,DiFilippo:2024mwm,Frolov:2024hhe,Bueno:2024zsx,Bueno:2025gjg,Aguayo:2025xfi,Hennigar:2025yqm,Xie:2025auj,Bueno:2025qjk,Bueno:2025tli,Hao:2025utc,Frolov:2026rcm,Tsuda:2026xjc,Li:2026mam,Bueno:2026dln,Sueto:2026epz,PinedoSoto:2026hfm} and~\cite{Borissova:2026wmn,Borissova:2026krh,Borissova:2026klg,Borissova:2026prd,Bueno:2025zaj,Konoplya:2026gim,Lutfuoglu:2026gis,Dubinsky:2026wcv,Malik:2026laq,Dubinsky:2026gcj,Saridakis:2026oyf,Colleaux:2017ibe}.}
 More generally, the equations of motion of generic $2$-dimensional Horndeski theories can be obtained as the reduced equations of motion on~\eqref{eq:MetricWarped} of purely gravitational  theories~\eqref{eq:S} involving non-polynomial functions of curvature and curvature-derivative invariants in $d\geq 4$ dimensions~\cite{Borissova:2026krh,Borissova:2026wmn,Colleaux:2019ckh}. Assuming a given $2$-dimensional  Horndeski theory satisfies the integrability condition~\eqref{eq:Integrability}, i.e., dealing with the warped-product sector~\eqref{eq:MetricWarped} of a general QTG, the generating function $\Omega$ can have an arbitrary dependence on $r$ and $f$ as independent variables~\cite{Borissova:2026krh}. \\

From a reversed perspective, the previous discussion implies the following remarkably general statement. 
Consider a static black hole~\eqref{eq:Metricf} asymptotically of the form~\eqref{eq:fGR} with an invertible dependence of $f$ on the mass $M$.
 It follows that any such black hole
  can be reconstructed as a vacuum solution to a $d$-dimensional generally covariant gravitational theory~\cite{Borissova:2026krh}. To that end, one may identify the black hole mass $M$ with the integration constant  in the integrated equation of motion~\eqref{eq:Omega2M} of an associated $2$-dimensional  Horndeski theory satisfying the integrability condition~\eqref{eq:Integrability}. This identification determines the generating function $\Omega$ offshell after performing the replacement $r \mapsto \varphi$ and $f \mapsto \chi$. Therefrom the functions $h_i$ can be reconstructed which make the configuration
\ba
q_{ab} \qty(y)\dd{y}^a \dd{y}^b  &=& - f(r)\dd{t}^2 + \frac{\dd{r}^2}{f(r)},\label{eq:qf}\quad \,\,\,\\
\varphi(y) &=& r\label{eq:Varphi},
\ea
a solution to the equations of motion for $q_{ab}$ and $\varphi$ derived from~\eqref{eq:L2D}~\cite{Boyanov:2025pes,Carballo-Rubio:2025ntd}. One may then interpret this Horndeski theory as the effective $2$-dimensional theory obtained from the reduction on~\eqref{eq:MetricWarped} of a general $d$-dimensional QTG~\eqref{eq:S}~\cite{Borissova:2026krh}.~\footnote{Such a reconstruction procedure has been developed  originally in the context of a unified extension of the Einstein equations on the space of warped-product spacetimes~\eqref{eq:MetricWarped} based on a second-order covariantly conserved rank-$2$ tensor constructed from the variations of the $2$-dimensional Horndeski action~\cite{Carballo-Rubio:2025ntd}, cf.~also~\cite{Boyanov:2025pes}. For a generic $2$-dimensional metric $q_{ab}$ and scalar field $\varphi$, the tensor constructed in~\cite{Carballo-Rubio:2025ntd} can be generated as the variation of a $d$-dimensional covariant action for $g_{\mu\nu}$ evaluated on~\eqref{eq:MetricWarped},~concretely equation~\eqref{eq:Emunu}~\cite{Borissova:2026krh}. Thus the extended Einstein equations~\cite{Carballo-Rubio:2025ntd} in vacuum are in fact equivalent to the reduced field equations of a $d$-dimensional covariant theory for $g_{\mu\nu}$~\cite{Borissova:2026krh}.} 

A concrete realisation of the functions $h_i$ in terms of the generating function $\Omega$ is provided by~\cite{Borissova:2026dlz}
\ba
  h_2(\varphi,\chi) &=& \frac{16 \pi G_\text{N} }{(d-2)\Sigma_{d-2}} \partial_\varphi \Omega(\varphi,\chi),\label{eq:h2}\\
 h_3(\varphi,\chi) &=& -\frac{32 \pi G_\text{N} }{(d-2)\Sigma_{d-2}}  \partial_\chi \Omega(\varphi,\chi),\label{eq:h3}\\
 h_4(\varphi,\chi) &=&  -\frac{16 \pi G_\text{N} }{(d-2)\Sigma_{d-2}}  \int \dd{\varphi} \partial_\chi \Omega(\varphi,\chi) .\label{eq:h4}\quad 
\ea
The goal here is to use the above results for a unified discussion of black hole thermodynamics based on the effective $2$-dimensional reduced theory~\eqref{eq:SHorndeski} without knowledge of its $d$-dimensional generally covariant form~\eqref{eq:S}.\\

{\it Thermodynamics of $g_{tt}g_{rr}=-1$ black holes---}Throughout we will assume that $f$ has a most outer zero at $r=r_+$ corresponding to the event horizon of the black hole. From now on we will only deal with functions $\mathcal{F}(r,f)$ evaluated at $r = r_+$ and hence $f=0$, and for this purpose adopt the short hand notion $\mathcal{F}\qty(r_+) \vcentcolon = \mathcal{F}\qty(r_+,0)$.
The integrated equation of motion~\eqref{eq:Omega2M} implies that an onshell configuration $f=f(r)$ corresponds to a level curve of $\Omega$ and therefore the mass $M$
can be written as
\ba\label{eq:MOmega}
M &=&
\Omega\qty(r_+).
\ea
We proceed by expressing the Hawking temperature $T$ in terms of $\Omega$. The former can be derived by an application of Euclidean
 path integral 
methods~\cite{Gibbons:1976ue,Hawking:1978jz}. The Euclidean metric obtained after a Wick rotation of the time coordinate $t \to \imath \tau$
 expanded at $r=r_+$ has a conical singularity unless $\tau$ is periodically identified with period $\kappa/2\pi$, where $\kappa = f'\qty(r_+)/2 > 0$ is the surface gravity computed from the timelike Killing vector field $\xi^\mu = \partial_t^\mu$ according to $\xi^\nu \nabla_\nu \xi^\mu = \kappa \xi^\mu$. The imaginary time period fixes the inverse of the Hawking temperature
 \ba\label{eq:T}
 T &=& \frac{f'\qty(r_+)}{4 \pi}.\quad 
 \ea
For a black hole solution to~\eqref{eq:Omega2M} with generating function $\Omega$, we can use~\eqref{eq:OmegaEOM} to obtain 
\ba\label{eq:TOmega}
T& =&  -\frac{1}{4\pi}\frac{\partial_{r_+} \Omega(r_+)}{\partial_f \Omega(r_+)}.\quad 
\ea
Now we will evaluate the black hole entropy by an application of Wald's Noether charge formalism. \\

{\it Wald entropy---}The defining formula for the Wald entropy $S$ of a black hole derived from a generally covariant action~\eqref{eq:S} is~\cite{Wald:1993nt,Iyer:1994ys}
 \ba\label{eq:SWald}
 S &=& - 2\pi  \oint_{\mathcal{H}} \dd[d-2]{x}\sqrt{h}\, \qty[\frac{P^{\mu\nu\rho\sigma}}{16 \pi G_\text{N}}] \epsilon_{\mu\nu}\epsilon_{\rho\sigma},\quad 
 \ea
where $h$ is the induced metric on the horizon $\mathcal{H}$ and $\epsilon_{\mu\nu}$ the binormal to $\mathcal{H}$ normalised to  $\epsilon_{\mu\nu}\epsilon^{\mu\nu}=-2$. The entropy tensor $P$ is defined as the variational derivative of the Lagrangian with respect to the Riemann tensor, 
 \ba\label{eq:PTens}
 P^{\mu\nu\rho\sigma} &=& \frac{\var \mathcal{L}_g}{\var R_{\mu\nu\rho\sigma}}.\quad 
 \ea
Its contraction with the binormals in~\eqref{eq:SWald} must be evaluated onshell.
 For a static black hole~\eqref{eq:Metricf} the horizon binormal is given in terms of the unit spacelike and timelike normal vectors $r^\mu $ and $t^\nu$ as $\epsilon_{\mu\nu} = 2 r_{[\mu} t_{\nu]} $ and has the non-zero components $\epsilon_{01} = -\epsilon_{10} = 1$. The contraction of the entropy tensor with the binormals in this case is constant on the horizon and the Wald formula~\eqref{eq:SWald} simplifies to
 \ba\label{eq:SWaldIntermediate}
 S &=& - \frac{ \Sigma_{d-2}}{8 G_\text{N}} \,r_+^{d-2} \eval{ P^{abcd}\epsilon_{ab}\epsilon_{cd}}_{r=r_+}.\quad 
 \ea
 The key step  now is to notice that realising such a black hole as a solution to a general QTG, the remaining tensorial contraction can be computed from the $2$-dimensional reduced Lagrangian density 
 \ba
 \eval{\mathcal{L}_g}_{\eqref{eq:MetricWarped}} &=& (d-2) \,\varphi^{2-d} \mathcal{L}_{q,\varphi}\quad 
 \ea
 defined as in~\eqref{eq:L2D}
  without knowledge of its $d$-dimensional generally covariant form.~\footnote{Analogue techniques for computing the Wald entropy in polynomial and non-polynomial curvature QTGs and more general gravitational theories admitting solutions of the form~\eqref{eq:Metricf} have been used e.g.~in~\cite{Colleaux:2019ckh,Bueno:2019ycr,Bueno:2022res,Bueno:2024dgm,Hennigar:2025yqm,Aguayo:2025xfi}.} The former depends only linearly on
\ba
\eval{R_{abcd}}_{\eqref{eq:MetricWarped}} &=& \mathcal{R}\, q_{a[c} q_{d]b}\quad 
\ea
via the $2$-dimensional  Ricci scalar in~\eqref{eq:L2D}, and in particular does not depend on covariant derivatives of the curvature. Thus the variational derivative in~\eqref{eq:SWaldIntermediate} reduces to a partial derivative. The relevant part of the reduced Lagrangian density is
  \ba
   \eval{\mathcal{L}_g}_{\eqref{eq:MetricWarped}} & \ni & (d-2) \,\varphi^{2-d} h_4(\varphi,\chi) \mathcal{R}.\quad 
  \ea
  Using $q^{a[c} q^{d]b}\epsilon_{ab}\epsilon_{cd} = 2 q^{-1}$ where the determinant of $q_{ab}$ is $q = -1$ for the ansatz~\eqref{eq:qf},
the Wald entropy evaluates to~\footnote{When the $2$-dimensional reduced theory~\eqref{eq:SHorndeski} features a linear non-minimal coupling $h_4(\varphi,\chi) \propto \varphi$, such as for instance in Jackiw-Teitelboim gravity~\cite{Jackiw:1984je,Teitelboim:1983ux}, equation~\eqref{eq:SWaldRed} reduces to the well-known statement that the entropy is proportional to the value of the dilaton field $\varphi$ on the horizon. See e.g.~\cite{Mertens:2022irh} for a review.}
\ba
S &=& \frac{ \Sigma_{d-2}}{4 G_{\text{N}}}\,r_+^{d-2}\, \qty[\eval{\partial_{\mathcal{R}}  \eval{\mathcal{L}_g}_{\eqref{eq:MetricWarped}}}_{\text{\eqref{eq:qf}--\eqref{eq:Varphi}}}]_{r=r_+}\quad\nn\\
&=& \frac{(d-2) \Sigma_{d-2}}{4 G_\text{N}}  \,h_4(r_+).\label{eq:SWaldRed}\quad 
\ea
The function $h_4(r_+)$ in~\eqref{eq:SWaldRed} still needs to be expressed onshell for a solution. When the black hole is realised through the representation of the function $h_4$ as in~\eqref{eq:h4}, we can write~\footnote{One may verify that this expression reproduces the entropy of asymptotically flat or anti-de Sitter black holes in $d\geq 5$ polynomial curvature QTGs~\cite{Bueno:2024dgm,Hennigar:2025yqm,Aguayo:2025xfi}. Changing variables $(r,f) \mapsto (r,\psi)$, where $\psi(r,f) = (k-f)/r^2$ with $k = \pm 1$, and inserting~\eqref{eq:OmegaPolynomialCQTG} into~\eqref{eq:SWaldRedOmega}, results in
	\ba
	S &=&
	- \frac{(d-2)\Sigma_{d-2} k^{d/2-1}}{8 G_{\text{N}}} \int \dd{\psi_+} \frac{h'\qty(\psi_+)}{\psi_+^{d/2}},\quad
	\ea
	where $\psi_+ = \psi\qty(r_+)= k/r_+^2$.}
\ba\label{eq:SWaldRedOmega}
S &=& -4 \pi  \int \dd{r_+} \partial_f \Omega\qty(r_+).\quad\,\,\,
\ea
Using the expressions for the mass $M$ in~\eqref{eq:MOmega} and temperature $T$ in~\eqref{eq:TOmega}, this result can be written as 
\ba\label{eq:SFirstLaw}
S &=& \int \dd{r_+} \frac{\partial_{r_+} M}{T},\quad
\ea
wherefrom the first law of thermodynamics 
\ba\label{eq:FirstLaw}
\var{M} &=& T \var{S}\quad
\ea
follows. In particular the entropy formula~\eqref{eq:SWaldRedOmega} combined with the integrated equation of motion~\eqref{eq:Omega2M} provide all ingredients required for a consistent discussion of the thermodynamics of $d$-dimensional static black holes~\eqref{eq:Metricf}. \\

{\it Generalised first law---}The above derivation assumes the thermodynamic mass to be a function of $r_+$ only, cf.~equation~\eqref{eq:MOmega}. More generally, when the black hole mass is used to determine the generating function $\Omega$ through~\eqref{eq:Omega2M} --- the key step to make this spacetime a solution ---
$\Omega$ evaluated at $r=r_+$ and hence $M$ will be
a function
\ba\label{eq:OmegaFunc}
M&=& M\qty(r_+,\Lambda ,\alpha_i)\quad 
\ea
with a dependence on all dimensionful scales involved in $f$, such as $\Lambda$ and other mass scales $\alpha_i$.
In this case one may consider a
 generalisation of the
first law~\eqref{eq:FirstLaw}
in the form
\ba\label{eq:FirstLawGeneralised}
\delta M  &=& T \var S + V \delta P + \sum_{i} \mu_{\alpha_i}\delta \alpha_i ,\quad 
\ea
where
\ba\label{eq:P}
P &=& - \frac{\Lambda}{8 \pi G_{\text{N}}} \quad \,\,\,
\ea
is a thermodynamic pressure associated to the cosmological constant~\cite{Teitelboim:1985dp,Brown:1988kg,Padmanabhan:2002sha}
with conjugate variable $V$ given by
\ba\label{eq:VDef}
V &=&  \qty(\pdv{M}{P})_{S,\alpha_i}
\ea
and  interpreted as thermodynamic volume of the black hole~\cite{Caldarelli:1999xj,Papadimitriou:2005ii,Kastor:2009wy,Dolan:2010ha,Cvetic:2010jb,Dolan:2013ft}, and
$\mu_{\alpha_i}$ are generalised thermodynamic potentials conjugate to the couplings $\alpha_i$.
Interpreting their contribution in~\eqref{eq:FirstLawGeneralised} as work terms, the black hole mass $M$ must be identified with the chemical enthalpy $H = E + PV$ rather the internal energy $E$ as in~\eqref{eq:FirstLaw} --- an observation which has critically sparked the field of black hole chemistry~\cite{Kubiznak:2014zwa,Kubiznak:2016qmn,Mann:2025xrb}.~\footnote{A holographic perspective on black hole chemistry and associated  interpretation of $P$ as a central charge can be found in~\cite{Karch:2015rpa,Visser:2021eqk,Zeyuan:2021uol,Ahmed:2023snm,Mann:2024sru}. A perspective challenging the view of $\Lambda$ as a thermodynamic variable is provided in~\cite{Mancilla:2024spp}. See also~\cite{Borsboom:2026ash} for a proposal of a holographic definition of thermodynamic pressure and volume for black holes.} The generalised first law~\eqref{eq:FirstLawGeneralised} in the absence of work terms associated with $\alpha_i$ was discovered originally as a way to construct a generalisation of the Smarr mass formula~\cite{Smarr:1972kt} for an anti-de Sitter Schwarzschild black hole~\eqref{eq:fGR}~\cite{Caldarelli:1999xj,Kastor:2009wy},
\ba\label{eq:Smarr}
(d-3)M &=&  (d-2)TS -2VP,\quad 
\ea
for which the additional $V \var{P}$ term in~\eqref{eq:FirstLawGeneralised} compared to~\eqref{eq:FirstLaw} is essential.~\footnote{Generalisations of the Smarr formula~\eqref{eq:Smarr} for higher-curvature gravitational theories with dimensionful couplings  have been constructed via the introduction of chemical potentials for Lovelock theories e.g.~in~\cite{Kastor:2010gq,Cai:2013qga,Henriquez-Baez:2022bfi} and for more general polynomial curvature QTGs e.g.~in~\cite{Hennigar:2015esa,Hennigar:2025yqm,Aguayo:2025xfi}.} One way to see this is to observe that the expression for the Wald entropy~\eqref{eq:SWaldRedOmega} is invariant under a shift of the generating function $\Omega$ by a function of $r$,
\ba
\Omega(r,f) & \to & \Omega(r,f) + \omega(r),\quad 
\ea
whereas the mass $M$ in~\eqref{eq:MOmega} is not. Such a shift modifies the solution to the equation of motion~\eqref{eq:Omega2M}. The cosmological constant term in~\eqref{eq:H1} is an example of such an $r$-dependent shift of $\Omega$,~\footnote{More generally, the addition of a scalar potential $V(\varphi)$ encoded in the function $h_2$ in~\eqref{eq:L2D} amounts to such an $r$-dependent shift of $\Omega$, cf.~the defining relations of $\Omega$ in~\eqref{eq:AlphaBeta1}--\eqref{eq:AlphaBeta2} based on the functions $\alpha$ and $\beta$ in~\eqref{eq:Alpha}--\eqref{eq:Beta}. This suggests that one might also deal with black holes exhibiting more general non-flat asymptotics than anti-de Sitter which would require an extended notion of black hole volume.}
\ba\label{eq:OmegaLambda}
\Omega_\Lambda(r) 
& = & \frac{(d-2)\Sigma_{d-2}}{16 \pi G_{\text{N}}} \qty[\frac{-2 \Lambda}{(d-1)(d-2)}] r^{d-1}.\quad \,\,\,
\ea
Upon performing the identification~\eqref{eq:Omega2M}, the thermodynamic volume $V$ in~\eqref{eq:VDef} is therefrom computed as
\ba\label{eq:V}
V &=&  \frac{\Sigma_{d-2}}{d-1} \,r_+^{d-1}.\quad 
\ea
A generalisation of the Smarr formula~\eqref{eq:Smarr} to take into account the new dimensionful parameters $\alpha_i$ can be derived by an application of
Euler's theorem for a homogenous function
\ba\label{eq:MHomogenous}
M\qty(\lambda r_+, \lambda^p \Lambda, \lambda^{q_i} \alpha_i) &=& \lambda^{s}M\qty(r_+, \Lambda, \alpha_i),\quad\,\,\,
\ea
which yields the scaling relation
\ba\label{eq:EulerScaling}
s M\qty(r_+, \Lambda, \alpha_i) &=& \qty(\pdv{M}{r_+})_{\Lambda,\alpha_i} r_+ + p \qty(\pdv{M}{\Lambda})_{r_+,\alpha_i} \Lambda \quad\,\,\,\,\, \nn\\
& +& \sum_i q_i  \qty(\pdv{M}{\alpha_i})_{r_+,\Lambda,\alpha_{j\neq i}}\alpha_i.
\ea 
The assumption of homogeneity of $M$ appears restrictive from the point of view of a unified picture of $d$-dimensional black hole thermodynamics on the space of general QTGs considered here, but can be achieved under quite generic circumstances.
For example, let us assume that the function $\Omega$ specifying the integrated equation of motion~\eqref{eq:Omega2M} of the $2$-dimensional reduced theory can be expanded as a series of terms beyond GR, in the form
\ba
\Omega(r,f) &=&  \frac{(d-2)\Sigma_{d-2}}{16 \pi G_{\text{N}}}r^{d-1}h(r,f),\label{eq:OmegaAnsatz}\quad \quad\quad  \\
h(r,f)&=& \frac{1}{\ell^2} + \frac{k-f}{r^2} + \sum_{i} \alpha_i \eta_i(r,f),\quad \label{eq:hAnsatz}
\ea
where $h$ can have a general dependence on $r$ and $f$ through the functions $\eta_i$.~\footnote{For polynomial curvature QTGs, cf.~equation~\eqref{eq:OmegaPolynomialCQTG}, $h$ is a power series in the variable $\psi =(k-f)/r^2$, i.e., $\eta_i \propto \psi^i$ for $i \geq 2$. This variable represents one of the scalar invariants characterising the Riemann tensor of the spacetimes~\eqref{eq:Metricf}. Such a scalar structure combining $r$ and $f$ can be separated so as to achieve a dependence of $h$ on $r$ and $f$ as independent variables by allowing covariant derivatives apart from curvatures in the action~\eqref{eq:S}~\cite{Borissova:2026krh}.} These functions have mass dimensions $\qty[\eta_i] = 2 - \qty[\alpha_i]$ and are
assumed to not depend on additional dimensionful variables beyond $r$. Hence they must be of the form
\ba
\eta_i(r,f) &=& \hat{\eta}_i(f) \,r^{\qty[\alpha_i]-2}.\quad 
\ea
The mass $M$ in~\eqref{eq:MOmega} can thus be expressed as
\ba\label{eq:MSeries}
M\qty(r_+,\Lambda,\alpha_i) &=&   \frac{(d-2)\Sigma_{d-2}}{16 \pi G_{\text{N}}}r_+^{d-1}\bigg[ \frac{1}{\ell^2(\Lambda)} + \frac{k}{r_+^2}\quad \quad \nn\\
&+ &\sum_i {\alpha_i} \hat{\eta}_i(0) r_+^{\qty[\alpha_i]-2}\bigg],
\ea
and
 is a homogenous function with scaling dimension $s=d-3$, whereas the scaling dimensions of $\Lambda$ and $\alpha_i$ are $p = -2$ and $q_i = - \qty[\alpha_i]$. In this case the Euler relation~\eqref{eq:EulerScaling} yields
\ba\label{eq:SmarrPreliminary}
(d-3)M &=& \qty(\pdv{M}{r_+})_{\Lambda,\alpha_i}r_+ - 2 VP\quad \quad\quad \,\,\,  \nn\\
& -& \sum_i \qty[\alpha_i] \qty(\pdv{M}{\alpha_i})_{r_+,\Lambda,\alpha_{j\neq i}} \alpha_i.\quad 
\ea
The entropy can be obtained according to~\eqref{eq:SWaldRedOmega} with $\Omega$ given in~\eqref{eq:OmegaAnsatz} as
\ba
S &=&
\frac{A}{4 G_{\text{N}}}\qty[1 -(d-2) \sum_i  \frac{\alpha_i  \hat{\eta}'_i(0)}{d-2+\qty[\alpha_i]} r_+^{ \qty[\alpha_i]}],\quad \,\,\,\,
\ea
where $A = \Sigma_{d-2}r_+^{d-2}$ is the $d-2$ dimensional area of the horizon. Using this result to compute
\ba
\qty(\pdv{M}{r_+})_{\Lambda,\alpha_i}r_+ &=& T \qty(\pdv{S}{r_+})_{\Lambda,\alpha_i}r_+,\,\,\,
\ea
one may construct a generalised Smarr formula
\ba
(d-3)M &=& (d-2)TS - 2 VP - \sum_{i}\qty[\alpha_i] \mu_{\alpha_i}\alpha_i\quad \,\,\,\,\,
\ea
for suitably defined theory-dependent potentials $\mu_{\alpha_i}$.
\\

{\it Global charges---}An electric charge $Q$ of the black hole originating from the coupling to an electromagnetic field which recovers Maxwell theory for $r\to \infty$ is manifested in an asymptotic $r$-dependent falloff term $\propto r^{2(3-d)}$ in $f$ which is subleading compared to the mass term. Upon performing the identification~\eqref{eq:Omega2M} with $M$ taken to be the black hole mass, such a global charge contribution would amount to a universal shift of $\Omega$ of the form~\eqref{eq:OmegaFunc} in the limit $r\to \infty$ but would require a model-dependent discussion of the conjugate potential $\Phi$ when considered as arising from the coupling of~\eqref{eq:S} to an electromagnetic source. Here we have generated the black hole as an effective vacuum solution through~\eqref{eq:Omega2M} such that a global charge parameter $Q$ should be viewed as just another mass scale similar to $\Lambda$ and $\alpha_i$. A proper way to treat charged black holes in this framework is to first generate the corresponding black hole with $Q=0$ as a vacuum solution through~\eqref{eq:Omega2M} and subsequently supplement the action~\eqref{eq:S} by the coupling to an electromagnetic source. The derivation of the Wald entropy~\eqref{eq:SWaldRedOmega} is insensitive to a  minimal coupling, whereas the solution to the equation of motion~\eqref{eq:Omega2M} is not. This would then be compensated by a work term $\Phi \var Q$ in~\eqref{eq:FirstLawGeneralised}. A generalisation of the framework presented here to include an electromagnetic field and is an apparent future extension.~\footnote{This could potentially broaden the perspective on near-extremal charged black holes beyond Einstein-Maxwell theory widely discussed based on $2$-dimensional Jackiw-Teitelboim gravity, see e.g.~\cite{Maldacena:2016upp,Almheiri:2016fws,Nayak:2018qej,Sachdev:2019bjn}.} 
\\

{\it Example---}Let us finish with an exemplary step-by-step application of the previous discussion. Consider the asymptotic anti-de Sitter regular Bardeen black hole in $d=4$~\cite{Bardeen:1968bh},
\ba\label{eq:fBardeen}
f(r) &=& -1 + \frac{r^2}{\ell^2} - \frac{2 m r^2}{[r^{2} + \gamma^{2}]^{\frac{3}{2}}},\quad 
\ea
where the parameter $m$ is related to the black hole mass $M$ by
\ba
M &=& \frac{\Sigma_{2}}{4 \pi G_{\text{N}}} \,m,\quad 
\ea
 as can be read off by taking the limit $r\to \infty$ of~\eqref{eq:fBardeen} and comparing with~\eqref{eq:fGR}. Moreover  $\gamma$ is a regularisation length parameter whose value we assume to be such that $f$ has a real zero and hence the metric describes a black hole. This metric belongs to a simple class of models for which $f$ is linear in $M$.
  Invoking the integrated equation of motion~\eqref{eq:Omega2M}, i.e., inverting~\eqref{eq:fBardeen} for $M$, the generating function $\Omega$ is identified as
 \ba
\Omega(r,f) &=& \frac{\Sigma_{2}}{8 \pi G_{\text{N}}\ell^2} \frac{1}{r^2} \qty[r^2 + \gamma^2]^{\frac{3}{2}} \qty[r^2 - (1+f)\ell^2].\,\,\,\quad 
 \ea
The thermodynamic mass $M$ in~\eqref{eq:MOmega}, temperature $T$ in~\eqref{eq:TOmega} and entropy $S$ in~\eqref{eq:SWaldRedOmega}
 are then computed as
\ba
M &=&  \frac{\Sigma_{2}}{8 \pi G_{\text{N}}\ell^2} \frac{1}{r_+^2} \qty[r_+^2 + \gamma^2]^{\frac{3}{2}} \qty[r_+^2 - \ell^2],\quad \,\,\,\\
T &=& \frac{1 }{4 \pi \ell^2 } \frac{1}{r_+}\frac{1}{r_+^2 + \gamma^2}\qty[3 r_+^4 - \ell^2 r_+^2 + 2 \gamma^2 \ell^2],\quad \,\,\, \\
S &=& \frac{\Sigma_{2}}{4 G_{\text{N}}r_+}\Bigg[ \qty[r_+^2 - 2 \gamma^2]\sqrt{ r_+^2+ \gamma^2}
\nn\\
&+ & 3 \gamma^2 r_+\, \text{arctanh} \qty[\frac{r_+}{\sqrt{r_+^2 + \gamma^2}}]\Bigg],
\ea
wherefrom the first law~\eqref{eq:FirstLaw} can be verified. In the limit $\gamma \to 0$ the entropy reduces to the Schwarzschild-anti-de Sitter area law $A/4G_{\text{N}}$ where $A = \Sigma_2 r_+^2$.\\

{\it Final remarks---}The entropy formula~\eqref{eq:SWaldRedOmega} establishes explicitly the corrections to the Bekenstein-Hawking area law for $d$-dimensional static and asymptotically flat or anti-de Sitter beyond-GR black holes satisfying $g_{tt}g_{rr}=-1$ in Schwarzschild gauge when regarded as vacuum solutions to a general QTG.
It will be insightful to supplement the derivation of the Wald entropy by a computation of the entropy based on Euclidean path integral methods~\cite{Gibbons:1976ue,Hawking:1978jz} to see whether these are equivalent or not~\cite{Iyer:1995kg}. Even though the boundary terms of the $d$-dimensional theory are generally unknown, one could consider an analysis at the level of the $2$-dimensional reduced theory using the boundary terms for Horndeski theory~\cite{Padilla:2012ze}. For static black holes these would be a generalised Gibbons-Hawking-York term involving the non-minimal coupling $h_4$ and supplemented by appropriate counterterms depending on the background asymptotics. It is overall fascinating to anticipate the depth of possibilities of addressing open questions in black hole physics~\cite{Buoninfante:2024oxl,Buoninfante:2024yth,Carballo-Rubio:2025fnc} based on {\it generic} $2$-dimensional dilaton theory as reduced $d$-dimensional gravity.

\begin{acknowledgments}
	
I thank Ra{\'u}l Carballo-Rubio and Jo\~ao Magueijo for discussions. This work is supported by STFC Consolidated Grant ST/X000575/1.
	
\end{acknowledgments}

\enlargethispage{20pt}

\bibliographystyle{jhep}
\bibliography{references}

\end{document}